\documentclass[hyper]{JHEP} 

\usepackage{epsfig}

\newcommand\fverb{\setbox\pippobox=\hbox\bgroup\verb}

\newcommand\fverbdo{\egroup\medskip\noindent%

            \fbox{\unhbox\pippobox}\ }

\newcommand\fverbit{\egroup\item[\fbox{\unhbox\pippobox}]}

\newbox\pippobox

\title{Is Bimetric Gravity Really  Ghost Free? }
\author{J. Kluso\v{n}\\
Department of
Theoretical Physics and Astrophysics\\
Faculty of Science, Masaryk University\\
Kotl\'{a}\v{r}sk\'{a} 2, 611 37, Brno\\
Czech Republic\\
E-mail: \email{klu@physics.muni.cz}}

 \abstract{We perform the Hamiltonian analysis of the
 bimetric theory of gravity introduced in
 [arXiv:1109.3515 [hep-th]]. We carefully analyze the requirement
 of the preservation of all constraints and
 we find that there is no additional constraint
 that could eliminate the ghost mode.}

\def\mR{\mathcal{R}}
\newcommand{\bC}{\mathbf{C}}
\newcommand{\bD}{\mathbf{D}}
\def\hf{\hat{f}}
\def\tK{\tilde{K}}
\def\tn{\tilde{n}}
\def\bA{\mathbf{A}}
\def\tmR{\tilde{\mR}}
\def\bF{\mathbf{F}}
\def\bV{\mathbf{V}}
\def\bW{\mathbf{W}}

\def\tnabla{\tilde{\nabla}}
\def\tx{\tilde{x}}
\def\tmG{\tilde{\mG}}
\def\mC{\mathcal{C}}
\def\be{\begin{equation}}
\def\bD{\mathbf{D}}
\def\ee{\end{equation}}
\def\bD{\mathbf{D}}
\def\bea{\begin{eqnarray}}

\def\eea{\end{eqnarray}}

\def\tr{\mathrm{tr}\, }

\def\bz{\mathbf{z}}

\def\tr{\mathrm{Tr}}

\def\bx{\mathbf{x}}
\def\by{\mathbf{y}}
\def \mD{\mathcal{D}}

\newcommand{\hg}{\hat{g}}

\newcommand{\mU}{\mathcal{U}}

\newcommand{\tD}{\tilde{D}}
\newcommand{\mG}{\mathcal{G}}

\def\mV{\mathcal{V}}

\def \bA{\mathbf{A}}

\newcommand{\bT}{\mathbf{T}}

\newcommand{\mL}{\mathcal{L}}

\def\pb #1{\left\{#1\right\}}

\begin{document}
\section{Introduction and Summary}
Recently  the new very interesting formulation  of the non-linear
massive gravity \cite{deRham:2010ik,deRham:2010kj} was introduced
with significant improvement reached in
\cite{Hassan:2011vm,Hassan:2011hr}. This theory was further extended
in \cite{Hassan:2011tf} where the theory was formulated with general
reference metric. The most crucial fact that is related to given
theory is the proof of the absence of the ghosts that are generally
expected in any theory that breaks the diffeomorphism invariance. As
is well know the physical content is determined in the Hamiltonian
formulation when all constraints are  identified together with their
nature.  This analysis was performed in  several papers
\cite{Kluson:2011qe,Kluson:2011aq,Golovnev:2011aa,Kluson:2011rt,Kluson:2012gz,Kluson:2012wf,Kluson:2012zz}
with the most important results derived in
\cite{Hassan:2011ea,Hassan:2012qv}  with the outcome that this
non-linear massive theory possesses one additional constraint and
the resulting constraint structure is sufficient for the elimination
of the ghost degree of freedom.

Very interesting extension of given theory was suggested  in
\cite{Hassan:2011zd} when the kinetic term for the general reference
metric was introduced and hence $\hg_{\mu\nu}$ and $\hf_{\mu\nu}$
come in the symmetric way in the action. Then it was argued
in \cite{Hassan:2011ea}
 that the resulting
theory is the ghost free formulation of the bimetric theory of
gravity \footnote{For further analysis of given theory, see
\cite{Hassan:2011zd,Hassan:2012rq,Nojiri:2012re,Soloviev:2012wr,
Kluson:2012ps,Hassan:2012gz,Hassan:2012wr,Nojiri:2012zu,
Nomura:2012xr,Baccetti:2012re,Berg:2012kn,Baccetti:2012bk,Hassan:2012wt,Hinterbichler:2012cn,Paulos:2012xe}.}.

The goal of this paper is to perform the Hamiltonian analysis of the
bimetric theory of gravity in the form introduced in
\cite{Hassan:2011zd}. For simplicity we call this theory as the new
bimetric theory of gravity (NBTG).
 We would like to explicitly determine the
structure of the constraints and eventually to prove the absence of
the ghosts. Remarkably we find very subtle issue related to NBTG
which forces us to doubt whether the ghost could be eliminated in
given theory or not.  More explicitly, the non-linear massive
gravity with general reference metric has the potential that depends
on the matrix $H^\mu_{ \ \nu}\equiv \hg^{\mu\rho}\hf_{\rho\sigma}$
where $\hf_{\rho\sigma}$ is \emph{fixed} background metric. There is
now no doubt that such theory is ghost free. On the other hand  the
situation changes in case of the bimetric theory of gravity when we
promote $\hf_{\mu\nu}$ as an additional dynamical field with the
kinetic term given by Einstein-Hilbert action. Then, since the
interaction term between two metrics has the square structure form
we  follow \cite{Hassan:2011ea,Hassan:2012qv,Hassan:2011zd} and
perform the redefinition of one shift function that makes the theory
linear in $N$ and $M$ which are the lapse functions in
$\hg_{\mu\nu}$ and $\hf_{\mu\nu}$ respectively. It is important that
the action has the same structure as the non-linear massive gravity
action with general reference metric $\hf_{\mu\nu}$ with additional
kinetic term for $\hf_{\mu\nu}$. However the fact that
$\hf_{\mu\nu}$ is dynamical has crucial impact on the Hamiltonian
structure of given theory. Explicitly, the Hamiltonian is given as
the linear combination of the constraints as opposite to the case of
the non-linear massive gravity where the Hamiltonian does not vanish
on the constraint surface.  Now the crucial point is that the
components of the metric $\hf_{\mu\nu}$ that appear as the
\emph{fixed} parameters in the non-linear massive gravity case
should be now considered as \emph{Lagrange multiplicators} whose
values are determined by the requirements of the preservation of all
constraints during the  time evolution of the system. However then
we find that the requirement of the preservation of the constraint
$\mC_0$ whose explicit definition will be given below leads to the
differential equation for the Lagrangian multiplicator $M$ that is
related to the $\mD$. In other words the value of the Lagrange
multiplicator $M$ is determined by the requirement of the
preservation of $\mC_0$ during the time evolution of the system. In
the same way we fix the value of the Lagrange multiplicator $N$. Say
differently, $\mC_0$ and $\mD$ could be interpreted as the second
class constraints.

 At this place we should compare our result with the known
proof of the absence of the ghosts in the bimetric theory. It was
shown  in the very nice paper \cite{Hassan:2011ea} that the
requirement of the preservation of the constraint $\mC_0$ implies an
additional constraint $\mC_{(2)}$ (in their notation) given in the
e.q. (3.32) in this paper. We see that this constraint contains the
covariant derivative of $M$ which, as we argued above, is
\emph{fixed} in case of the non-linear massive gravity theory so
that it is really natural to interpret $\mC_{(2)}$ as an additional
constraint. Then this constraint together with $\mC_0$ are
responsible for the elimination of the ghost mode which is crucial
for the consistency of non-linear massive gravity. However in case
of the bimetric gravity $M$  should be considered as the Lagrange
multiplicator whose value is fixed by the consistency of given
theory.

The fact that $\mC_0$ and $\mD$ should  be interpreted as the second
class constraints has important consequence for the dynamics of the
theory. More precisely,  since  the Hamiltonian is given as the
linear combination of the constraints implies  that the resulting
Hamiltonian vanishes strongly. This is rather puzzling result and we
believe that this is a consequence of the redefinition of the shift
function \cite{Hassan:2011ea,Hassan:2012qv,Hassan:2011zd} which is
certainly useful  for the non-linear massive gravity where the
diffeomorphism invariance is explicitly broken but we are not sure
whether it is suitable for the bimetric theory of gravity where all
fields are dynamical and the theory is manifestly diffeomorphism
invariant under diagonal diffeomorphism.
 In fact, it
is non-trivial task to identify such generator as was shown recently
in \cite{Kluson:2012ps} in case of particular model of bimetric
theory of gravity \cite{Damour:2002ws}. We are currently analyzing
NBTG following \cite{Kluson:2012ps} and we believe that it is
possible to identify four first class constraints that are
generators of the diagonal diffeomorphism. On the other hand the
analysis performed  so far suggests that  it is very difficult or
even impossible to find an additional constraint that could
eliminate the additional mode \footnote{This analysis will appear in
forthcoming publication.}.

We should however stress that we have to be very careful with
definitive conclusions.  We wanted to show that the
 extension of the non-linear massive
 gravity to the bimetric theory of gravity as was performed in
\cite{Hassan:2011zd} could be  more subtle that we initially thought
and that it is not really clear whether given theory is ghost free
or not.  It is still possible that there  exists
 the way how to find four first class constraints that are generators
 of diagonal
diffeomorphism together with additional constraints that eliminate
ghost mode. Clearly more work is needed in order to resolve this
issue.


This paper is organized as follows. In the next section
(\ref{second}) we review the Hamiltonian formulation of the bimetric
theory of gravity \cite{Hassan:2011zd} and identify primary and the
secondary constraints. Then in section (\ref{third}) we calculate
the algebra of constraints for the case of the minimal version of
the bimetric gravity. Finally in    Appendix \ref{apendix} we
briefly discuss  the  Hamiltonian formulation of the bimetric $F(R)$
theory of gravity.

\section{Bimetric Gravity}\label{second}
In this section we review the main properties of the bimetric theory
of gravity in the formulation presented in \cite{Hassan:2011zd}. The
starting point is following action
\begin{eqnarray}
S&=&M_g^2\int d^4x \sqrt{-\hg}{}^{(4)}R^{(g)}+M_f^2 \int d^4x
\sqrt{-\hf}{}^{(4)} R^{(f)}+
\nonumber \\
&+&2m^2 M_{eff}^2\int d^4x\sqrt{-\hg}\sum_{n=0}^4 \beta_n e_n
(\sqrt{-\hg \hf}) \ , \nonumber \\
\end{eqnarray}
where \begin{equation} M_{eff}^2= (\frac{1}{M_g^2}+
\frac{1}{M_f^2})^{-1} \  ,
\end{equation}
and  where $\hg_{\mu\nu},\hf_{\mu\nu}$ are four-dimensional metric
components with ${}^{(4)}R^{(g)}, {}^{(4)}R^{(f)}$ corresponding
scalar curvatures. Further,
 $e_k(\bA)$ are elementary
symmetric polynomials of the
eigenvalues of $\bA$. For generic
$4\times 4$ matrix they are given by
\begin{eqnarray}
e_0(\bA)&=& 1 \ , \nonumber \\
e_1(\bA)&=&[\bA] \ , \nonumber \\
e_2(\bA)&=&\frac{1}{2}([\bA]^2-[\bA^2]) \
, \nonumber \\
e_3(\bA)&=&\frac{1}{6} \left( [\bA]^3-3[\bA][\bA^2]+2[\bA^3]\right)
\
, \nonumber \\
e_4(\bA)&=&\frac{1}{24} \left([\bA]^4-6[\bA]^2[\bA^2]+3
[\bA^2]^2+8[\bA][\bA^3]-6[\bA^4]\right)
\ , \nonumber \\
e_k(\bA)&=&0 \ , \mathrm{for} \ k>4 \ ,
\end{eqnarray}
where $\bA^\mu_{ \ \nu}$ is $4\times 4$
matrix and where
\begin{equation}
[\bA]=\tr \bA=\bA^\mu_{ \ \mu} \ .
\end{equation}
Of the four $\beta_n$ two combinations are related to the mass and
the cosmological constant while the remaining two combinations are
free parameters. If we consider the case when the cosmological
constant is zero and the parameter $m$ is mass, the four $\beta_n$
are parameterized in terms of the $\alpha_3$ and $\alpha_4$ of
\cite{deRham:2010ik,deRham:2010kj}
\begin{equation}
\beta_n=(-1)^n\left(\frac{1}{2}(4-n)(3-n)-
(4-n)\alpha_3+\alpha_4\right) \ .
\end{equation}
The minimal action corresponds to $\beta_2=\beta_3=0$ that implies
$\alpha_3=\alpha_4=1$ and consequently $\beta_0=3 \ , \beta_1=-1$.

Our goal is to find the Hamiltonian formulation of given theory and
determine corresponding primary and the secondary constraints. As
the first step we introduce $3+1$ decomposition
of
 both $\hg_{\mu\nu}$ and
$\hf_{\mu\nu}$ \cite{Gourgoulhon:2007ue,Arnowitt:1962hi}
\begin{eqnarray}
\hat{g}_{00}&=&-N^2+N_i g^{ij}N_j \ , \quad \hat{g}_{0i}=N_i \ ,
\quad \hat{g}_{ij}=g_{ij} \ ,
\nonumber \\
\hat{g}^{00}&=&-\frac{1}{N^2} \ , \quad \hat{g}^{0i}=\frac{N^i}{N^2}
\ , \quad \hat{g}^{ij}=g^{ij}-\frac{N^i N^j}{N^2} \
\nonumber \\
\end{eqnarray}
and
\begin{eqnarray}
\hf_{00}&=&-M^2+L_i f^{ij}L_j \ , \quad \hf_{0i}=L_i \ , \quad
\hf_{ij}=f_{ij} \ , \nonumber \\
\hf^{00}&=&-\frac{1}{M^2} \ , \quad \hf^{0i}=-\frac{L^i}{M^2} \ ,
\quad \hf^{ij}=
f^{ij}-\frac{L^i L^j}{M^2} \ , \quad  L^i=L_jf^{ji} \ , \nonumber \\
\end{eqnarray}
and where we defined $g^{ij}$ and  $f^{ij}$ as the inverse to
$g_{ij}$ and $f_{ij}$ respectively
\begin{equation}
g_{ik}g^{kj}=\delta_i^{ \ j} \ , \quad  f_{ik}f^{kj}=\delta_i^{ \ j}
\ .
\end{equation}
Following
\cite{Hassan:2011vm,Hassan:2011hr,Hassan:2011tf,Hassan:2011zd} we
  perform following redefinition of the shift function
\begin{equation}
N^i=M\tn^i+L^i+N \tD^i_{ \ j } \tn^j \
\end{equation}
so that the resulting action is linear in $M$ and $N$. Note that the
matrix $\tD^i_{ \ j}$ obeys the equation \cite{Hassan:2011vm,Hassan:2011hr,Hassan:2011tf,Hassan:2011zd}
\begin{equation}\label{defD}
\sqrt{\tx}\tD^i_{ \ j}= \sqrt{(g^{ik}-\tD^i_{ \ m} \tn^m \tD^k_{ \
n}\tn^n)f_{kj}}
\end{equation}
and also following important property
\begin{eqnarray}
f_{ik}\tD^k_{ \ j}
= f_{jk}\tD^k_{ \ i } \ .  \nonumber \\
\end{eqnarray}
Then after some calculations we derive the bimetric gravity action
in the form
\cite{Hassan:2011vm,Hassan:2011hr,Hassan:2011tf,Hassan:2011zd}
\begin{eqnarray}\label{SRbi}
S&=&M_f^2\int dt d^3 \bx  M\sqrt{f}[\tK_{ij}\tmG^{ijkl}\tK_{kl}+
 R^{(f)}]+ M_g^2\int dt d^3\bx
N\sqrt{g}[K_{ij}\mG^{ijkl}K_{kl}+R^{(g)}
]+ \nonumber \\
&+& 2m^2M_{eff}^2 \int  dt d^3\bx \sqrt{g} (M \mU+N\mV) \ ,
 \nonumber \\
\end{eqnarray}
where
\begin{eqnarray}
K_{ij}&=&\frac{1}{2N}(\partial_t g_{ij}- \nabla_i
N_j(\tn,g)-\nabla_j
N_i(\tn,g)) \ , \nonumber \\
 \tK_{ij}&=&\frac{1}{2M}(\partial_t f_{ij}- \tnabla_i
L_j-\tnabla_j L_i) \ , \nonumber \\
\end{eqnarray}
where
\begin{equation}
N_i=M g_{ij}\tn^j+g_{ij}L^j+Ng_{ik}\tD^k_{ \ j}\tn^j \ , \quad
L_i=f_{ij}L^j \ ,
\end{equation}
and  where $\nabla_i, R^{(g)}$ and $\tnabla_i,R^{(f)}$ are the
covariant derivatives and scalar curvatures  calculated using
$g_{ij}$ and $f_{ij}$ respectively. Further,
 $\mG^{ijkl}$ and $\tmG^{ijkl}$ are
de Witt metrics defined as
\begin{equation}
\mG^{ijkl}=\frac{1}{2}(g^{ik}g^{jl}+g^{il}g^{jk})-g^{ij}g^{kl} \ ,
\quad  \tmG^{ijkl}=\frac{1}{2}(f^{ik}f^{jl}+f^{il}f^{jk})-
f^{ij}f^{kl} \
\end{equation}
with inverse
\begin{equation}
\mG_{ijkl}=\frac{1}{2}(g_{ik}g_{jl}+ g_{il}g_{jk})-\frac{1}{2}
g_{ij}g_{kl} \ , \quad \tmG_{ijkl}=\frac{1}{2}(f_{ik}f_{jl}+
f_{il}f_{jk})-\frac{1}{2} f_{ij}f_{kl} \
\end{equation}
that obey the relation
\begin{equation}
\mG_{ijkl}\mG^{klmn}=\frac{1}{2}(\delta_i^m\delta_j^n+
\delta_i^n\delta_j^m)  \ , \quad
\tmG_{ijkl}\tmG^{klmn}=\frac{1}{2}(\delta_i^m\delta_j^n+
\delta_i^n\delta_j^m)  \ .
\end{equation}
Finally, $ \mV$ and $\mU$ introduced  in (\ref{SRbi}) have the form
\begin{eqnarray}
\mV&=&\beta_0 +\beta_1 \sqrt{\tx}\tD^i_{ \
i}+\beta_2\frac{1}{2}\sqrt{\tx}^2(\tD^i_{ \ i}\tD^j_{ \ j}- \tD^i_{
\
j} \tD^j_{ \ i})+ \nonumber \\
&+&\frac{1}{6}\beta_3  \sqrt{\tx}^3[ \tD^i_{ \ i} \tD^j_{ \
j}\tD^k_{ \ k}-3\tD^i_{ \ i}\tD^j_{ \ k} \tD^k_{ \ j} +2\tD^i_{ \
j}\tD^j_{ \ k}\tD^k_{ \ i}] \ ,
\nonumber \\
\mU&=&\beta_1 \sqrt{\tx}+ \beta_2  [\sqrt{\tx}^2 \tD^i_{ \ i}+\tn^i
f_{ij}\tD^j_{ \ k}\tn^k]+ \nonumber
\\
&+&\beta_3[\sqrt{\tx}(\tD^l_{ \ l}\tn^i f_{ij} \tD^j_{ \ k}\tn^k-
\tD^i_{ \ k}\tn^k f_{ij}\tD^j_{ \ l}\tn^l)+ \frac{1}{2} \sqrt{\tx}^3
(\tD^i_{ \ i}\tD^j_{ \ j}-\tD^i_{ \ j}\tD^j_{ \ i})]
 + \beta_4 \frac{\sqrt{f}}{\sqrt{g}} \ ,  \nonumber \\
\end{eqnarray}
where
\begin{equation}
\tx=1-\tn^if_{ij}\tn^j \ .
\end{equation}
 The action (\ref{SRbi}) is suitable for the Hamiltonian formalism. First we
find the momenta conjugate to $N,\tn^i$ and $g_{ij}$
\begin{equation}
\pi_N\approx 0 \ , \quad  \pi_i\approx 0 \ , \quad
\pi^{ij}=M_g^2\sqrt{g}\mG^{ijkl}K_{kl} \
\end{equation}
together with the momenta conjugate to $M,L^i$ and $f_{ij}$
\begin{equation}
\rho_M\approx 0 \ , \quad \rho_i\approx 0 \ , \quad \rho^{ij}= M_f^2
\sqrt{f} \tmG^{ijkl}\tK_{kl} \ .
\end{equation}
Then after some calculations we find following Hamiltonian
\begin{eqnarray}
H&=&\int d^3\bx (\pi^{ij}\partial_t g_{ij}+ \rho^{ij}
\partial_t f_{ij}-\mL)=
\nonumber \\
&=&\int d^3\bx (N\mC_0+M \mD+L^i \mR_i) \ , \nonumber \\
\end{eqnarray}
where
\begin{eqnarray}
\mC_0&=& \frac{1}{M_g^2\sqrt{g}} \pi^{ij}
\mG_{ijkl}\pi^{kl}-M_g^2\sqrt{g} R^{(g)}+\mR^{(g)}_k\tD^k_{
\ l}\tn^l -2m^2M_{eff}^2\sqrt{g}\mV \ ,  \nonumber \\
\mD&=& \frac{1}{M_f^2\sqrt{f}} \rho^{ij}\tmG_{ijkl}
\rho^{kl}-M_f^2\sqrt{f}R^{(f)}+\tn^i\mR_i^{(g)}-2m^2M_{eff}^2\sqrt{g} \mU  \ ,  \nonumber \\
\mR_i&=&\mR_i^{(g)}+\mR_i^{(f)} \ , \nonumber \\
\end{eqnarray}
where we also denoted
\begin{equation}
\mR^{(g)}_i=-2g_{ik}\nabla_l\pi^{lk} \ , \quad  \mR^{(f)}_i=
-2f_{ik} \tnabla_l\rho^{lk} \ .
\end{equation}
From previous analysis we see that we have eight primary constraints
\begin{equation}
\pi_N\approx 0 \ , \quad \pi_i \approx 0 \ , \quad  \rho_M\approx  0 \ ,
\quad \rho_i\approx 0 \ .
\end{equation}
Then the next step is to analyze the requirement that these
constraints are preserved during the time evolution of the system
\begin{eqnarray}
\partial_t\pi_N&=&\pb{\pi_N,H}=-\mC_0\approx  0 \ , \nonumber \\
\partial_t \rho_M&=&\pb{\rho_M,H}=-\mD \approx 0 \ , \nonumber \\
\partial_t\pi_i&=&\pb{\pi_i,H}=
 \mC_k \left(M\delta^k_i+N\frac{\delta (\tD^k_{ \ j}\tn^j)}{\delta
\tn^i}\right) \approx 0  \ , \nonumber \\
\partial_t\rho_i&=&\pb{\rho_i,H}= -\mR_i \approx 0 \ ,  \nonumber \\
\end{eqnarray}
where
\begin{eqnarray}
\mC_i&=&\mR_i^{(g)}+2m^2M_{eff}^2\sqrt{g} \frac{\tn^p
f_{pm}}{\sqrt{\tx}} [\beta_1 \delta^m_{ \ i}+ \beta_2 [\delta^m_{ \
i}\tD^l_{ \ l}- \tD^m_{ \
i}]+\nonumber \\
&+&\beta_3  \sqrt{\tx}^2 (\frac{1}{2}\delta^m_{ \ i}(\tD^n_{ \
n}\tD^p_{ \ p}-\tD^m_{ \ n} \tD^n_{ \ m})+\tD^m_{ \ l}\tD^l_{ \
i}-\tD^m_{ \ i}\tD^n_{ \ n}) ]
\ , \nonumber \\
\end{eqnarray}
and where we used  the canonical Poisson brackets
\begin{eqnarray}
\pb{N(\bx),\pi_N(\by)}&=&\delta(\bx-\by) \ , \quad
 \pb{\tn^i(\bx),\pi_j(\by)}=\delta^i_{j}\delta(\bx-\by) \ , \nonumber \\
\pb{g_{ij}(\bx),\pi^{kl}(\by)}&=& \frac{1}{2} (\delta_i^k\delta_j^l+
\delta_i^l\delta_j^k)\delta(\bx-\by) \ , \nonumber \\
\pb{M(\bx),\rho_M(\by)}&=&\delta(\bx-\by) \ , \quad
\pb{L^i(\bx),\rho_j(\by)}=\delta^i_{j}\delta(\bx-\by) \ ,
\nonumber \\
 \pb{f_{ij}(\bx),\rho^{kl}(\by)}&=& \frac{1}{2}
(\delta_i^k\delta_j^l+\delta_i^l\delta_j^k)\delta(\bx-\by) \
\nonumber \\
\end{eqnarray}
and also following important relations \cite{Hassan:2011ea}
\begin{eqnarray}
\frac{\delta \sqrt{\tx}\tD^i_{ \ i}}{\delta \tn^i}
&=&-\frac{1}{\sqrt{\tx}}\tn^n f_{nm}\frac{\delta (\tD^m_{ \
p}\tn^p)}
{\delta \tn^i} \ ,  \nonumber \\
\frac{\partial}{\partial \tn^i}\tr (\sqrt{\tx}\tD)^2 &=&
-2 \tn^p f_{pm} \tD^m_{ \ k} \frac{\delta (\tD^k_{ \
n}\tn^n)}{\delta \tn^i} \ ,
\nonumber \\
\frac{\delta}{\delta \tn^i} \tr (\sqrt{\tx}\tD)^3&=&
 -3\sqrt{\tx} \tn^kf_{km} \tD^m_{ \
n} \tD^n_{ \ p} \frac{\delta (\tD^p_{ \ n}\tn^n)} {\delta \tn^i} \ .
 \nonumber \\
\end{eqnarray}
In summary we have following  $16$ constraints
\begin{eqnarray}\label{const}
\mathrm{primary}: \  \pi_N\approx 0 \ , \pi_i\approx 0 \ ,
\rho_M\approx 0 \ ,
\rho_i\approx 0 \ ,  \nonumber \\
\mathrm{secondary:} \ \mC_0\approx 0 \ , \mD\approx 0 \ , \mC_i
\approx 0 \ , \mR_i\approx 0 \ .\nonumber \\
\end{eqnarray}
Now we have to check the stability of all constraints when the total
Hamiltonian takes the form
\begin{eqnarray}
H_T&=&\int d^3 \bx (N\mC_0+M\mD+L^i\mR_i+u^N\pi_N+
u^i\pi_i+ \nonumber \\
&+&v^M\rho_M+v^i\rho_i +\Sigma^i\mC_i) \ ,  \nonumber \\
\end{eqnarray}
where $N,M,L^i,u^N,u^i,v^M,v^i,\Sigma^i$ are Lagrange multiplicators
related to the constraints (\ref{const}).
 For simplicity we restrict ourselves to the case of the minimal
bi-metric theory.
\section{Preservation of Constraints in Case of Minimal Bimetric Gravity}\label{third}
The  minimal bimetric theory is defined by
 the following choice
of parameters
\begin{equation}
\beta_0=3 \ , \beta_1=-1 \ , \beta_2=0 \ , \beta_3=0 \ , \beta_4=1 \
.
\end{equation}
Now we proceed to the analysis of the preservation of all
constraints given in (\ref{const}).  It is easy to see
that  the constraint $\pi_N\approx 0$ is trivial preserved. On the
other hand  the requirement of the preservation  of the constraint
$\pi_i\approx 0$ takes the form
\begin{eqnarray}\label{parttpii}
\partial_t\pi_i=\pb{\pi_i,H_T}=-\left(M\delta_{i}^{k}+
\frac{\partial (\tD^k_{ \ j}\tn^j)}{\partial \tn^i} \right) \mC_k+
\int d^3\bx \Sigma^j\pb{\pi_i,\mC_j(\bx)} =0 \ ,
\nonumber \\
\end{eqnarray}
where
\begin{eqnarray}
 \pb{\pi_i(\bx),\mC_j(\by)}=
 \left[\frac{1}{\sqrt{\tx}}f_{ij}+ \frac{\tn^kf_{ki}
\tn^lf_{lj}}{\sqrt{\tx}^3}\right]\delta(\bx-\by) \equiv
\triangle_{\pi_i,\mC_j}\delta(\bx-\by)
 \ .
\nonumber \\
\end{eqnarray}
Since
\begin{eqnarray}\label{dettriangle}
\det \triangle_{\pi_i,\mC_j}=\det
\left(\frac{f_{ik}}{\sqrt{\tx}}\right) \det \left(\delta^k_{ \
j}+\frac{1}{\tx}\tn^k\tn^mf_{mj}\right)=
\frac{1}{\tx^{5/2}} \det f_{ij} \neq 0
 \nonumber \\
\end{eqnarray}
we find that   $\triangle_{\pi_i,\mC_j}$ is non-singular matrix on
the whole phase space. However this fact also implies that the
equation (\ref{parttpii}) has trivial solution
\begin{equation}
\Sigma^i=0 \ .
\end{equation}
In the samy we proceed with the analysis of the time evolution of
the constraint $\mC_i$
\begin{eqnarray}\label{partmC}
\partial_t\mC_i(\bx)&=&\pb{\mC_i(\bx),H_T}=
\int d^3\by \left(N(\by)\pb{\mC_i(\bx),\mC_0(\by)}+ \right.
\nonumber \\
& &\left.+M(\by)\pb{\mC_i(\bx),\mD(\by)}+ v^j(\by)
\pb{\mC_i(\bx),\pi_j(\by)}\right)=0 \ . \nonumber \\
\end{eqnarray}
According to  (\ref{dettriangle}) we see that (\ref{partmC}) can be
solved for $v^i$ as functions of the canonical variables and $N,M$.
Say differently, $\pi_i$ and $\mC_i$ are the second class
constraints.

As the next step we consider the constraint $\mR_i$. It turns out
that is convenient to extend it by the expression $
\partial_i \tn^j\pi_j+
\partial_j(\tn^j\pi_i)$
and consider its smeared form
\begin{equation}
\bT_S(N^i)= \int d^3\bx N^i(\mR_i^{(g)}+\mR_i^{(f)}+p_\phi
\partial_i\phi+\partial_i \tn^j\pi_j+
\partial_j(\tn^j\pi_i))\equiv \int d^3\bx N^i\tmR_i \ .
\end{equation}
Then using the canonical Poisson brackets we  find
\begin{eqnarray}\label{bTgf}
\pb{\bT_S(N^i),g_{ij}(\bx)}&=&-\partial_k g_{ij}(\bx)N^k(\bx)-
\partial_i N^k(\bx)g_{kj}(\bx)-g_{ik}(\bx)\partial_j N^k(\bx) \ ,
\nonumber \\
\pb{\bT_S(N^i),\pi^{ij}(\bx)}&=&-\partial_k (N^k(\bx)
\pi^{ij}(\bx))+\partial_k N^i(\bx)\pi^{kj}(\bx)+
\pi^{ik}(\bx)\partial_k N^j(\bx) \ , \nonumber \\
\pb{\bT_S(N^i),f_{ij}(\bx)}&=&-\partial_k f_{ij}(\bx)N^k(\bx)-
\partial_i N^k(\bx)f_{kj}(\bx)-f_{ik}(\bx)\partial_j N^k(\bx) \ ,
\nonumber \\
\pb{\bT_S(N^i),\rho^{ij}(\bx)}&=&-\partial_k (N^k(\bx)
\rho^{ij}(\bx))+\partial_k N^i(\bx)\rho^{kj}(\bx)+
\rho^{ik}(\bx)\partial_k N^j(\bx) \ , \nonumber \\
\pb{\bT_S(N^i),\tn^i(\bx)}&=&-N^k(\bx)\partial_k\tn^i(\bx)+
\partial_j N^i(\bx)\tn^j(\bx) \ ,
 \nonumber \\
\pb{\bT_S(N^i),\pi_i(\bx)}&=&-\partial_k(N^k\pi_i)(\bx)
-\partial_i N^k(\bx)\pi_k(\bx) \ . \nonumber \\
\end{eqnarray}
Then we easily find the familiar result
\begin{equation}
\pb{\bT_S(N^i),\bT_S(M^j)}= \bT_S((N^j\partial_j M^i- M^j
\partial_j N^i)) \ .
\end{equation}
To proceed further we need to know the Poisson bracket between
$\bT_S(N^i)$ and $\tD^i_{ \ j}$ which can be determined
when we know the  explicit form
of $\tD^i_{ \ j}$
\cite{Hassan:2011vm,Hassan:2011hr,Hassan:2011tf,Hassan:2011zd}
\begin{eqnarray}\label{tDiexp}
 \tD^i_{ \ j}=\sqrt{g^{ik}f_{km}Q^m_{ \ n}}(Q^{-1})^n_{ \ j} \ ,
 \nonumber \\
\end{eqnarray}
where
\begin{equation}\label{Qij}
Q^i_{ \ j}=\tx \delta^i_{ \ j}+\tn^i \tn^k f_{kj} \ , (Q^{-1})^j_{ \
k}=\frac{1}{\tx}(\delta^j_{ \ k}-\tn^j \tn^mf_{mk}) \ .
\end{equation}
Using the explicit form of $\tD^i_{ \ j}$ given in (\ref{tDiexp}) we
see that the Poisson bracket between $\bT_S(N^i)$ and $\tD^i_{ \ j}$
is determined by the Poisson brackets between $\bT_S(N^i)$ and $
g_{ij},f_{ij}$ and $Q^i_{ \ p}$. The Poisson brackets between
$\bT_S(N^i)$ and $g_{ij}$ and $f_{ij}$ were given in (\ref{bTgf})
and the Poisson bracket between $\bT_S(N^i)$ and $Q^i_{ \ j}$ can be
easily determined using (\ref{bTgf}) and (\ref{Qij})
\begin{eqnarray}
\pb{\bT_S(N^i),Q^i_{ \ j}}&=&-\partial_k Q^i_{ \ j}N^k+
\partial_k N^i \tn^k\tn^mf_{mj}-
\tn^i\tn^mf_{mk}\partial_j N^k= \nonumber \\
&=&-\partial_k Q^i_{ \ j}N^k+\partial_k N^iQ^k_{ \ j}- Q^i_{ \
k}\partial_j N^k \ .
\end{eqnarray}
Then with the help of this result we find
\begin{equation}
\pb{\bT_S(N^i),\tD^i_{ \ j}}= -\partial_k \tD^i_{ \ j}N^k
+\partial_k N^i\tD^k_{ \ j}-\tD^i_{ \ k}\partial_j N^k \
\end{equation}
and finally collecting all these results we  obtain
\begin{eqnarray}
\pb{\bT_S(N^i),\mC_0}&=&-\partial_i \mC_0 N^i-
\partial_i N^i\mC_0 \ , \nonumber \\
\pb{\bT_S(N^i),\mD}&=&-\partial_i \mD N^i-
\partial_i N^i\mD \ , \nonumber \\
\pb{\bT_S(N^i),\mC_i}&=&-\partial_j
N^j\mC_i-N^j\partial_j\mC_i-\partial_iN^j\mC_j \ .
\nonumber \\
\end{eqnarray}
Then it is easy to see  that $\bT_S(N^i)$ is preserved during the
time evolution of the system and that it corresponds to the
generator of the spatial diffeomorphism. In other words $\tmR_i$ are
 first class constraints.

Now we come to the calculation of the Poisson brackets between   the
constraints $\mC_0$ and $\mD$. It turns out that it is useful to
introduce the smeared form of these constraints
\begin{equation}
\bC(N)=\int d^3\bx N(\bx)\mC_0(\bx) \ , \quad \bD(M)= \int d^3\bx
M(\bx)\mD(\bx) \ .
\end{equation}
We begin with the Poisson bracket between $\mC_0(\bx)$ and
$\mC_0(\by)$. Since $\mC_0$ does not depend on $\rho^{ij}$ we
immediately find that the Poisson bracket between $\mC_0(\bx)$ and
$\mC_0(\by)$ has the same form as in \cite{Hassan:2011ea} which
means that  it vanishes on the constraint surface
\begin{equation}
\pb{\mC_0(\bx),\mC_0(\by)}\approx 0  \ .
\end{equation}
In case of $\mD$ we find
\begin{eqnarray}
& &\pb{\bD(M),\bD(N)}=\nonumber \\
&=&\int d^3\bx (\partial_i MN-\partial_iNM) [f^{ij}
(\mR_j^{(f)}+\mR_j^{(g)})+(\tn^i\tn^j-f^{ij})\mR_j^{(g)} +\tn^i 2m^2
M_{eff}^2\sqrt{g}\sqrt{\tx}]=\nonumber \\
&=& \int d^3\bx (\partial_i MN-\partial_iNM) [f^{ij}\mR_j +
(\tn^i\tn^j-f^{ij})\mC_j] \ . \nonumber \\
\end{eqnarray}
We see that the right side vanishes on the constraint surface.
Finally we come to the calculation of the Poisson bracket between
$\bC(M)$ and $\bD(N)$
\begin{eqnarray}\label{pbbCbD}
\pb{\bC(N),\bD(M)}
&=&-\int d^3\bx M\tn^i\partial_i N \mC_0+\int d^3\bx  NM\left(
\frac{4m^2
M_{eff}^2}{M_g^2\sqrt{g}}\pi^{ij}\mG_{ijkl}U^{kl}+\right. \nonumber \\
&+& \tD^j_{ \ m}\tn^m\partial_j\tn^i\mR_i^{(g)}- \tn^j\partial_j
(\tD^i_{ \ m}\tn^m)\mR_i^{(g)}  +2 \mR^{(g)}_k\frac{\delta (\tD^k_{
\ m}\tn^m)} { \delta
f_{ij}}\frac{1}{M_f^2\sqrt{f}}\mG_{ijkl}\rho^{kl}
\nonumber \\
&+&\left. 2m^2M_{eff}^2\tn^i\partial_i\mV +4
m^2M_{eff}^2\frac{\sqrt{g}}{\sqrt{f}M_f^2}
\tilde{V}^{mn}\tmG_{mnkl}\rho^{kl}\right)+\nonumber \\
&+&\int d^3\bx [N\tD^j_{ \ m}\tn^m\tn^i\mR^{(g)}_i\partial_j M-
M\tn^j\partial_j N\tD^i_{ \ m}\tn^m\mR_i^{(g)}] -
\nonumber \\
&-&4m^2 M_{eff}^2\int d^3\bx [NV^{kl}\nabla_l(M\tn^i)g_{ik}-
\nabla_p (N\tD^k_{ \ l}\tn^l)g_{km}U^{mp}M] \ ,  \nonumber
\\
\end{eqnarray}
 where
\begin{eqnarray}
U^{kl}=\frac{\delta (\sqrt{g}\mU)}{\delta g^{kl}} \ ,
\tilde{V}^{mn}= \frac{\delta \mV}{\delta f^{mn}} \ ,
V^{kl}=\sqrt{g}\frac{\delta \mV}{\delta g^{kl}} \ , \nonumber \\
\end{eqnarray}
Let us analyze the Poisson bracket calculated above in more details.
First of all we see that the first expression vanishes on the
constraint surface $\mC_0\approx 0$ which is desired result. On the
other hand in order to analyze the time evolution of the local
constraint $\mC_0$ it is useful to express the local form of the
Poisson bracket from  (\ref{pbbCbD}) that can be schematically
written as
\begin{eqnarray}\label{pbbCbDs}
& &\pb{\bC(N),\bD(M)}= \nonumber \\
&=& \int d^3\bz ( N(\bz)M(\bz) \bF(\bz)
 +\partial_{z^i} N(\bz)\bV^i(\bz)M(\bz)+
N(\bz)\partial_{z^i} M(\bz) \bW^i(\bz)) \ ,
\nonumber \\
\end{eqnarray}
where the explicit form of $\bF,\bV^i,\bW^i$ follow from
(\ref{pbbCbD}). Let us now  write
\begin{equation}
N(\bz)=\int d^3\bx N(\bx)\delta(\bx-\bz) \ , M(\bz)= \int d^3\by
M(\by)\delta(\by-\bz)
\end{equation}
and insert it to the right side  of the Poisson bracket
(\ref{pbbCbDs}). Then after some calculation we find that  it is
equal to
\begin{eqnarray}\label{pbbCbDNM1}
\int d^3\bx d^3\by N(\bx)M(\by)[\delta(\bx-\by)\bF(\bx)+
 \frac{\partial}{\partial y^i}
 \delta(\bx-\by)\bV^i(\by)+\frac{\partial}{\partial x^i}
 \delta(\bx-\by)\bW^i(\bx)] \ .  \nonumber \\
\end{eqnarray}
On the other hand  we have
\begin{equation}\label{pbbCbDNM}
\pb{\bC(N),\bD(M)}= \int d^3\bx d^3\by N(\bx)M(\by)
\pb{\mC(\bx),\mD(\by)} \ .
\end{equation}
Since (\ref{pbbCbDNM1}) and (\ref{pbbCbDNM}) have to match
for  any $N(\bx),M(\by)$ we
obtain
\begin{equation}
\pb{\mC(\bx),\mD(\by)}= \delta(\bx-\by)\bF(\bx)+
 \frac{\partial}{\partial y^i}
 \delta(\bx-\by)\bV^i(\by)+\frac{\partial}{\partial x^i}
 \delta(\bx-\by)\bW^i(\bx) \ .
\end{equation}
Now using this expression we can easily determine the requirement of
the preservation of the constraint $\mC_0$ during the time evolution
of the system
\begin{eqnarray}\label{coneq}
\partial_t \mC_0(\bx)&=&\pb{\mC_0(\bx),H_T}\approx
\int d^3\by M(\by)\pb{\mC(\bx),\mD(\by)}=\nonumber \\
 &=&M(\bx)[F(\bx)-\partial_i\bV^i(\bx)]+
 \frac{\partial M(\bx)}{\partial x^i}
 [W^i(\bx)-V^i(\bx)]=0 \nonumber \\
 \end{eqnarray}
This is the most crucial point of our calculation that deserves
careful explanation.  Let us imagine that we have $\bV^i=\bW^i$.
Then (\ref{coneq}) has solution   either $M(\bx)=0$ or $
F(\bx)-\partial_i \bV^i(\bx)=0$. In fact, the first case occurs when
the expression $F(\bx)-\partial_i \bV^i(\bx)$ is non-zero on the
whole phase space, as for example in case when this expression is
constant. On the other hand when  $F(\bx)-\partial_i \bV^i(\bx)$
depends on the  phase space variables  it is  more natural to impose
the condition $\mC_0^{(II)}\equiv F(\bx)-\partial_i \bV^i(\bx)=0$ as
an additional constraint. This would be the desired result since now
we would have two second class constraints $\mC_0,\mC_0^{II}$ that
would be sufficient for elimination of the ghost mode. Unfortunately
as we can see from (\ref{pbbCbD}) $\bV^i\neq \bW^i$ and the
situation is completely different since the equation
 (\ref{coneq}) cannot leave $M$ undetermined. Rather we should
 interpret (\ref{coneq}) as equation that can be solved for
 $M$ as
function of the phase space variables. In
 fact, in the same way we can analyze the requirement of the
 preservation of the constraint $\mD$ that again leads to the
 differential equation that can be solved for $N$.
In other words we mean that it is now natural to interpret
 $\mC_0$ together with $\mD$ as the second class constraints.
Certainly this is very strange result. In particular, now we find
that the total Hamiltonian strongly vanishes up the diffeomorphism
constraint. Of course, we know that this cannot be right since the
theory possesses the overall diffeomorphism invariance and hence
there should be four the first call constraints that are generators
of this diffeomorphism. The way how to find such  generators for
bimetric theory of gravity was suggested in \cite{Kluson:2012ps} at
least for particular bimetric gravity model. The extension of this
work to the case of the non-linear bimetric gravity is currently
under consideration. Then the result derived in this section
suggests that the redefinition of the shift function which is very
useful in the case of the non-linear massive gravity may not be the
right way in  the case of the bimetric theory of gravity.

Despite of the fact that the total Hamiltonian vanishes it is
instructive to count the number of the physical degrees of freedom.
Recall that phase space variables are
$N,\pi_N,\tn^i,\pi_i,M,\rho_M,L^i,\rho_i,f_{ij},\rho^{ij},
g_{ij},\pi^{ij}$ so that the total number of the phase space degrees
of freedom is $N_{p.s.d.f.}=40$. On the other hand we have
$N_{f.c.}=8$ first class constraints $\pi_N\approx 0 \ ,
\rho_M\approx 0 \ , \rho_i\approx 0 \ , \tmR_i\approx 0 $. Finally
we have $N_{s.c.}=8$ second class constraints $\mC_0\approx 0 \ ,
\mD\approx 0 \ , \pi_i\approx 0 \ , \mC_i\approx 0$. Then the number
of the physical degrees of freedom is equal to
\cite{Henneaux:1992ig}
\begin{equation}
N_{f.d.f.}=N_{p.s.d.f.}-2N_{f.c.}-N_{s.c.}=16 \ .
\end{equation}
At linearized level we can identify four degrees of freedom
corresponding to the massless graviton, ten degrees of freedom
corresponding to the massive graviton and two additional degrees of
freedom corresponding to the ghost mode. It is important to stress
that the same result can be found when we identify four first class
constraints corresponding to the diagonal diffeomorphism and also
additional eight second class constraints as in case of the bimetric
gravity model analyzed in \cite{Kluson:2012ps}. Of course, the
square root structure of the potential has remarkable property in
case of the non-linear massive gravity and maybe it could be useful
in case of the bimetric gravity as well. We only say that the step
from the non-linear massive gravity to the bimetric gravity is not
straightforward as it seems to be.
\appendix
\section{Hamiltonian Analysis of $F(R)$ Bimetric Gravity}\label{apendix}
In this appendix  we briefly perform the Hamiltonian formulation of
$F(R)$ bimetric theory of gravity which was introduced by S.Odintsov
and Nojiri in \cite{Nojiri:2012zu}.

 The
starting point is the action for the non-linear bimetric gravity
theory
\begin{eqnarray}\label{SactA}
S&=&M_g^2\int d^4x \sqrt{-\hg}R^{(g)}+M_f^2 \int d^4x \sqrt{-\hf}
R^{(f)}+
\nonumber \\
&+&2m^2 M_{eff}^2\int d^4x\sqrt{-\hg}\sum_{n=0}^4 \beta_n e_n
(\sqrt{-\hg \hf}) \ . \nonumber \\
\end{eqnarray}
Then in order to construct the $F(R)$ analogue of the bimetric
massive gravity we add following expression to the action
(\ref{SactA})
\begin{equation}
S_1=-M_g^2\int d^4x\sqrt{-\hg} \left(\frac{3}{2}\hg^{\mu\nu}
\partial_\mu\phi\partial_\nu\phi+V(\phi)\right) \ .
\end{equation}
Then with the help of the Weyl transformation
\begin{eqnarray}
\hg'_{\mu\nu}&=&e^\phi \hg_{\mu\nu} \ , \nonumber \\
R[\hg]&=&e^{\phi}(R[\hg']-\frac{3}{2} \hg'^{\mu\nu}\nabla'_\mu\phi
\nabla'_\nu\phi+3\hg'^{\mu\nu}\nabla'_\mu\nabla'_\nu\phi) \nonumber
\\
\end{eqnarray}
we find that  $S_{tot}=S_{bi}+S_1$ takes the form
\begin{eqnarray}\label{SFRbi}
S_{FR}&=&M_f^2\int d^4 x\sqrt{-\hf} R^{(f)}+M_g^2 \int d^4x
\sqrt{-\hg'}[e^{-\phi}R[\hg']-e^{-2\phi}V(\phi)]+ \nonumber \\
&+& 2m^2M_{eff}^2 \int d^4x \sqrt{-\hg'}\sum_{n=0}^4 \beta_n
e^{(\frac{n}{2}-2)\phi} e_n
(\sqrt{\hg'^{-1}\hf}) \nonumber \\
\end{eqnarray}
using
\begin{equation}
e_n(\sqrt{\hg^{-1}f})= e^{\phi/2} e_n(\sqrt{\hg'^{-1}\hf}) \ .
\end{equation}
In what follows we will consider (\ref{SFRbi}) as the definition of
the F(R) bimetric theory of gravity. Of course we should be able to
solve the equation of motion  for $\phi$ at least in principle so
that we could express $\phi$ as the function of $\hg',\hf'$. Then
inserting back to the action (\ref{SFRbi}) we obtain the action
 that is
non-linear function of $R[\hg']$ and hence has the form of the
$F(R)$ theory of gravity \footnote{For review, see
\cite{DeFelice:2010aj,Nojiri:2010wj}}. For that reason we can name
(\ref{SFRbi}) as bimetric $F(R)$ theory of gravity even if its
definition using the scalar field is more natural. Finally, in the
following we omit $'$ over $\hg_{\mu\nu},\hf_{\mu\nu}$.

To proceed further we perform the redefinition of the shift $N^i $
as in section (\ref{second}) so that we find the  $F(R)$ bigravity
action in the form
\begin{eqnarray}\label{SFRbia}
S_{FR}&=& M_f^2\int dt d^3 \bx
[M\sqrt{f}\tK_{ij}\tmG^{ijkl}\tK_{kl}+
\sqrt{f} MR_f]+\nonumber \\
&+& M_g^2 \int dt d^3\bx
\sqrt{g}N[e^{-\phi}K_{ij}\mG^{ijkl}K_{kl}+e^{-\phi}R^{(g)}
-e^{-2\phi}V(\phi)]+ \nonumber \\
&+& 2m^2M_{eff}^2 \int dt d^3\bx \sqrt{g}N (M \mU+N\mV) \ .
 \nonumber \\
\end{eqnarray}
Note that (\ref{SFRbia}) has similar form as the action (\ref{SRbi})
up to presence of the additional scalar potential $V(\phi)$ and
powers of the factor $e^\phi$. Explicitly, we have
%
\begin{eqnarray}
\mV&=&\beta_0 e^{-2\phi}+\beta_1
e^{-\frac{3}{2}\phi}\sqrt{\tx}\tD^i_{ \
i}+\beta_2e^{-\phi}\frac{1}{2}\sqrt{\tx}^2(\tD^i_{ \ i}\tD^j_{ \ j}-
\tD^i_{ \
j} \tD^j_{ \ i})+ \nonumber \\
&+&\frac{1}{6}\beta_3 e^{-\frac{1}{2}\phi} \sqrt{\tx}^3[ \tD^i_{ \
i} \tD^j_{ \ j}\tD^k_{ \ k}-3\tD^i_{ \ i}\tD^j_{ \ k} \tD^k_{ \ j}
+2\tD^i_{ \ j}\tD^j_{ \ k}\tD^k_{ \ i}] \ ,
\nonumber \\
\mU&=&\beta_1 e^{-\frac{3}{2}\phi}\sqrt{\tx}+ \beta_2 e^{-\phi}
[\sqrt{\tx}^2 \tD^i_{ \ i}+\tn^i f_{ij}\tD^j_{ \ k}\tn^k]+ \nonumber
\\
&+&\beta_3e^{-\frac{1}{2}\phi}[\sqrt{\tx}(\tD^l_{ \ l}\tn^i f_{ij}
\tD^j_{ \ k}\tn^k- \tD^i_{ \ k}\tn^k f_{ij}\tD^j_{ \ l}\tn^l)+
\frac{1}{2} \sqrt{\tx}^3 (\tD^i_{ \ i}\tD^j_{ \ j}-\tD^i_{ \
j}\tD^j_{ \ i})]
 + \beta_4 \frac{\sqrt{f}}{\sqrt{g}} \ .  \nonumber \\
\end{eqnarray}
Now using the action (\ref{SFRbia}) we can find the corresponding
Hamiltonian.  Firstly  we find the momenta conjugate to $N,\tn^i$
and $g_{ij}$
\begin{equation}
\pi_N\approx 0 \ , \quad \pi_i\approx 0 \ , \quad
\pi^{ij}=M_g^2\sqrt{g}e^{-\phi}\mG^{ijkl}K_{kl} \
\end{equation}
together with the momenta conjugate to $N,L_i$ and $f_{ij}$
\begin{equation}
\rho_M\approx 0 \ , \quad  \rho^i\approx 0 \ , \quad  \rho^{ij}=
M_f^2\sqrt{f} \tmG^{ijkl}\tK_{kl} \ .
\end{equation}
Since the action (\ref{SFRbi}) does not contain the time derivative
of $\phi$ we find that the momentum conjugate to $\phi$ is zero
 \begin{equation}
p_\phi\approx 0 \ . \end{equation}
 As a result we find following
Hamiltonian
\begin{eqnarray}
H&=&\int d^3\bx (\pi^{ij}\partial_t g_{ij}+ \rho^{ij}
\partial_t f_{ij}-\mL)=
\nonumber \\
&=&\int d^3\bx (N\mC_0+M \mD+L^i \mR_i) \ ,  \nonumber \\
\end{eqnarray}
where
\begin{eqnarray}
\mC_0&=&e^\phi \frac{1}{M_g^2\sqrt{g}} \pi^{ij}
\mG_{ijkl}\pi^{kl}-e^{-\phi}\sqrt{g}M_p^2
R^{(g)}+e^{-2\phi}\sqrt{g}M_p^2 V+\mR^{(g)}_k\tD^k_{
\ l}\tn^l -2m^2M_{eff}^2\sqrt{g}\mV \ ,  \nonumber \\
\mD&=&\frac{1}{\sqrt{f}M_f^2} \rho^{ij}\tmG_{ijkl}
\rho^{kl}-M_f^2\sqrt{f}R^{(f)}+\tn^i\mR_i^{(g)}-2m^2M_{eff}^2\sqrt{g} \mU \ ,  \nonumber \\
\mR&=&\mR_i^{(f)}+\mR_i^{(g)} \ . \nonumber \\
\end{eqnarray}
Comparing with the situation in the second section we see that there
is an additional primary constraint $p_\phi\approx 0$. Then again
the requirement of the preservation of the primary constraints
implies the secondary constraints that have the same form as in case
of pure bimetric theory of gravity. There is however an additional
constraint $\mG$ that follows from the requirement of the
preservation of the constraint $p_\phi\approx 0$
\begin{eqnarray}
\partial_t p_\phi&=&\pb{p_\phi,H}=N\left(-\frac{e^\phi}{M_g^2\sqrt{g}}
\pi^{ij}\mG_{ijkl}\pi^{kl}+e^{-\phi}\sqrt{g}M_p^2 R^{(g)}-\right.
\nonumber
\\
 &-& \left. 2e^{-2\phi}\sqrt{g}M_p^2V-e^{-2\phi} \sqrt{g}M_p^2
\frac{dV}{d\phi}- 2m^2M_{eff}^2\sqrt{g}\frac{\delta \mV}{\delta
\phi}\right)\equiv -N\mG\approx 0 \ .
\nonumber \\
\end{eqnarray}
In summary we have following set  of $18$ constraints
\begin{eqnarray}
\mathrm{primary}: \  \pi_N\approx 0 \ , \pi_i\approx 0 \ ,
\rho_M\approx 0 \ ,
\rho_i\approx 0 \ , p_\phi\approx 0  \nonumber \\
\mathrm{secondary:} \mC_0\approx 0 \ , \mD\approx 0 \ , \mC_i
\approx 0 \ , \mR_i\approx 0 \ , \mG\approx 0 \ .\nonumber \\
\end{eqnarray}
Note that now the  constraint $\mC_i$ has  explicit form
\begin{eqnarray}
\mC_i&=&\mR_i^{(g)}+2m^2M_{eff}^2\sqrt{g}\times \nonumber \\
&\times &\frac{\tn^p f_{pm}}{\sqrt{\tx}} [\beta_1
e^{-3/2\phi}\delta^m_{ \ i}+ \beta_2 e^{-\phi}[\delta^m_{ \
i}\tD^l_{ \ l}- \tD^m_{ \
i}]+\nonumber \\
&+&\beta_3 e^{-\phi/2} \sqrt{\tx}^2 (\frac{1}{2}\delta^m_{ \
i}(\tD^n_{ \ n}\tD^p_{ \ p}-\tD^m_{ \ n} \tD^n_{ \ m})+\tD^m_{ \
l}\tD^l_{ \ i}-\tD^m_{ \ i}\tD^n_{ \ n}) ] \ .  \ \nonumber \\
\end{eqnarray}
Now we should check the stability of all constraints when the total
Hamiltonian takes the form
\begin{eqnarray}
H_T&=&\int d^3 \bx (N\mC_0+M\mD+L^i\mR_i+u_\phi p_\phi+u^N\pi_N+
u^i\pi_i+ \nonumber \\
&+& v^M\rho_M+v^i\rho_i +u_\phi^{II}\mG+\Sigma^i\mC_i) \ . \nonumber \\
\end{eqnarray}
It is easy to see that the Hamiltonian structure of $F(R)$ bimetric
theory of gravity is almost the same as the structure of NBTG
analyzed in previous two sections
 with small  exception that there are
two additional constraints $p_\phi\approx 0 \ , \mG\approx 0$. They
are the  second class constraints that vanish strongly and can be
explicitly  solved with respect to $p_\phi$ and $\phi$ at least in
principle. On the other hand they do  not affect the analysis of all
remaining constraints so that the constraint structure of given
theory is the same as in case of non-linear bimetric gravity. For
that reason we will not repeat the calculations performed in section
(\ref{third}).

 \noindent {\bf
Acknowledgements:}

 This work   was
supported  by the Grant agency of the Czech republic under the grant
P201/12/G028. \vskip 5mm

\end{document}